\documentclass[doublecol]{epl2}
\usepackage{amsmath}
\usepackage{amssymb}
\usepackage{color}

\DeclareMathSymbol{\lang}{\mathord}{symbols}{"68}
\DeclareMathSymbol{\rang}{\mathord}{symbols}{"69}
\DeclareMathSymbol{\openbra}{\mathord}{symbols}{"68}
\DeclareMathSymbol{\closeket}{\mathord}{symbols}{"69}

\newcommand{\ket}[1]{{| #1 \closeket}}

\title{Chaotic scattering of atoms at a standing laser wave}
\shorttitle{Chaotic scattering of atoms at a standing laser wave}

\author{S.\,V. Prants}
\shortauthor{S.\,V. Prants}

\institute{
Laboratory of Nonlinear Dynamical Systems,\\Pacific Oceanological
Institute of the Russian Academy of Sciences,\\43 Baltiiskaya st., 690041
Vladivostok, Russia, URL: dynalab.poi.dvo.ru}

\abstract{Atoms, propagating across a detuned standing laser wave,
can be scattered in a chaotic way even in the absence of spontaneous emission
and any modulation of the laser field.
Spontaneous emission masks the effect in some degree, but
the Monte Carlo simulation shows that it can be
observed in real experiments by the absorption imaging method or depositing
atoms on a substrate.
The effect of chaotic scattering is explained by a specific behavior of
the dipole moments of atoms crossing the field nodes and is shown to depend
strongly on the value of the atom-laser detuning.}

\begin{document}
\maketitle

\section{Introduction}
The deflection of an atomic beam at a laser standing wave (SW) is explained
by the dipole forces which are well described by the classical atom-field
interaction model \cite{Kazantsev,Arimondo}. The ability of a SW to diffract,
focus or splitting an atomic beam \cite{Mlynek}
has been used for a variety of applications including atom microscopy,
interferometry, isotope separation, and optical lithography
\cite{Timp,McClelland,Ober}. On the other hand, cold atoms are ideal
candidates to test fundamental principles of quantum physics including
the phenomenon of dynamical chaos at the microscopic level that is known
as a kind of random-like motion in a deterministic environment
\cite{GSZ92,Kolovsky,Steck01,R96,C09,Haake,Stockmann,JETPL97,JETPL01,PRE02}.
Dynamical chaos is characterized by exponential sensitivity of trajectories
in the phase space to small variations in initial conditions and/or control
parameters.

It has been proposed in Ref.~\cite{GSZ92} to study quantum chaos and the
corresponding effect of dynamical localization placing cold atoms in
a far-detuned SW with a periodic kick-like modulation of positions of
the SW nodes. A number of experiments \cite{Steck01} have
been carried out in accordance with this proposal. At large detunings,
the atoms are not excited being quantum analogues of classical kick rotors.
Since those experiments on atom optics realization of the
$\delta$-kicked quantum rotor, cold atoms provide new grounds
for experiments and theory on quantum chaos.

It has been shown in
Ref.~\cite{R96} that even a single-pulse far-detuned SW can induce chaos
in atomic motion. For sufficiently large detuning, the excited state amplitude
can be adiabatically eliminated~\cite{GSZ92}, leading to a Hamiltonian
with an external degree of freedom only. The corresponding equations of
motion for an externally modulated nonlinear pendulum constitute
the well-known model with one and half degree of freedom that can be chaotic
under some conditions. The other possibility is to induce chaos in
spin degree of freedom of atoms periodically kicked by applying short magnetic
field pulses \cite{C09}. That is the one and half degree of freedom model
of a kicked top.

In difference from those and other papers on the related topic,
we consider the physical situation with comparatively small detunings
and should take into account a coupling between
external and internal atomic dynamics, leading to
a model with three degrees of freedom. It will be shown in the present paper
that in this case one needs no modulation or any other perturbation of the SW
to induce chaotic internal and external dynamics of atoms crossing the SW
laser field.

Near the atom-field resonance, when
the interaction between the internal and external atomic degrees of freedom is
intense, there is a possibility to create conditions for chaotic
behavior without any kicking and modulation \cite{JETPL01,PRE02,PRA01}.
If so, it is open the way to test the novel regime of atomic motion caused
by the peculiarities of the dipole force in the strong coupling regime.
In the semiclassical approximation, atom with quantized internal dynamics is
treated as a point-like particle with the Hamilton--Schr\"odinger
equations of motion constituting a nonlinear dynamical system
\cite{JETPL01,PRE02,PRA01,PRA07}.
In a certain range of the atom-field detunings, a set of atomic trajectories
becomes exponentially sensitive to small variations in initial quantum
internal and classical external states or/and in the control parameters. Hamiltonian evolution
is a smooth process that is well described in a semiclassical approximation
by the Hamilton-Schr\"odinger equations. The problem becomes much more complicated
because of spontaneous emission of atoms with a specific shot quantum noise
acting in a dynamical system which is chaotic in the absence of noise.
A number  of nonlinear Hamiltonian and dissipative effects have been
found numerically and analytically near the resonance
including chaotic Rabi oscillations,
chaotic atomic transport, dynamical fractals, and
L\'evy flights \cite{JETPL01,PRE02,PRA01,PRA07,JETPL02,JETP03,PRA05,PRA08}.

\begin{figure}[htb]
\begin{center}
\includegraphics[width=0.3\textwidth,clip]{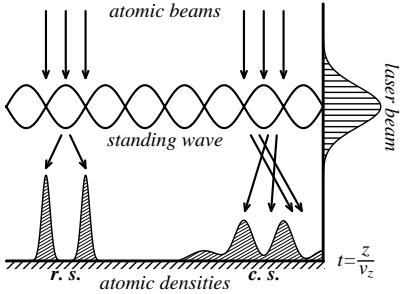}
\end{center}
\caption{Schematic representation of the proposed experiment on
regular (r.s.) and chaotic (c.s.) atomic scattering
at a Gaussian standing laser wave.}
\label{fig1}
\end{figure}

The main aim of the  paper is to demonstrate theoretically and numerically
that the new type of atomic diffraction
at a rigid SW without any modulation, {\it chaotic atomic scattering}, can be observed in
a real experiment. The scheme of such an experiment is shown in
Fig.~\ref{fig1} with a beam of atoms crossing a SW laser field.
One either measures a spatial atomic distribution after the
interaction by the absorption
imaging technique or measures an atomic distribution on a silicon substrate
in the far-field zone. The results are expected to be different depending on
the value of the atom-field detuning. The distribution is expected to be
comparatively narrow at those values of the detuning at which atomic
scattering is regular
(r.s. distribution in Fig.~\ref{fig1}) or wide at the detuning values providing
chaotic scattering of atoms due to their chaotic walking along the SW
(c.s. distribution in Fig.~\ref{fig1}).

\section{Results}
\subsection{Main equations and the regimes of atomic motion}

A beam of two-level atoms in the $z$ direction crosses
a SW laser field with optical axis in the $x$ direction (Fig.~\ref{fig1}).
The laser field \revision {amplitude} has the Gaussian profile
$\exp[-(z-z_0)^2/r^2]$ with $r$ being the $e^{-2}$ radius at the laser beam waist.
The characteristic length of the atom-field interaction is supposed to be
$3r$ because the light intensity at \revision {$z=z_0 \pm 1.5r$} is two orders of
magnitude smaller than the peak value. The longitudinal velocity of atoms,
$v_z$,
is much larger than their transversal velocity $v_x$ and is supposed to be
constant. Thus, the spatial laser profile may be replaced by the temporal one.
The Hamiltonian of an atom in the 1D SW
field can be written in the frame rotating with the laser frequency $\omega_f$
as follows:
\begin{equation}
\begin{gathered}
\hat H=\frac{P^2}{2m_a}+\frac{\hbar}{2}(\omega_a-\omega_f)\hat\sigma_z-\\
\hbar \Omega_0\exp[-(t-\frac{3}{2} \sigma_t)^2/\sigma^2_t]
\left(\hat\sigma_-+\hat\sigma_+\right)\cos{k_f X}
-\frac{i\hbar\Gamma}{2}\hat\sigma_+\hat\sigma_-,
\end{gathered}
\label{Ham}
\end{equation}
where $\hat\sigma_{\pm, z}$ are the Pauli operators for the internal atomic degrees
of freedom, $X$ and $P$ are the classical atomic position and momentum,
$\Gamma$, $\omega_a$, and $\Omega_0$ are the decay rate, the atomic
transition and maximal Rabi frequencies, respectively, and \revision {$\sigma_t\equiv
r/v_z$, i.e., $3\sigma_t$ is the transit time.}
The wavefunction for the electronic degree of freedom is
$\ket{\Psi(t)}=a(t)\ket{2}+b(t)\ket{1}$, where $a \equiv A+i\alpha$ and
$b\equiv B+i\beta$ are the probability amplitudes to find the
atom in the excited, $\ket{2}$, and ground, $\ket{1}$, states, respectively.
\begin{figure}[htb]
\begin{center}
\includegraphics[width=0.3\textwidth,clip]{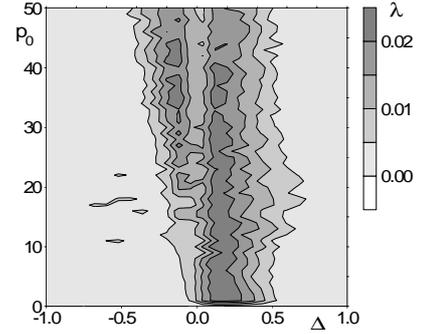}
\end{center}
\caption{\revision {Finite-time Lyapunov exponent} $\lambda$ vs atom-field detuning
$\Delta$ (in units of the Rabi frequency $\Omega_0$)
and initial atomic transversal momentum $p_0$ (in units of the photon momentum
$\hbar k_f$) at the normalized recoil frequency $\omega_r=10^{-3}$ and
$\gamma=0$.}
\label{fig2}
\end{figure}

In the semiclassical approximation, atom with quantized internal dynamics is
treated as a point-like particle \revision {(the transversal atomic momentum $p$ is
supposed to be, in average, much larger than the photon one},
$\hbar k_f$) with the
equations of motion written for the real and imaginary parts of the probability
amplitudes
\begin{equation}
\begin{gathered}
\dot x =\omega_r p, \, \dot p =- 2\exp[-(\tau-\frac{3}{2}
\sigma_{\tau})^2/\sigma^2_{\tau}](AB+\alpha \beta)\sin x,
\\
\dot A = \frac12 (\omega_rp^2 - \Delta)\alpha - \frac12 \gamma A
- \exp[-(\tau-\frac{3}{2} \sigma_{\tau})^2/\sigma^2_{\tau}]\beta\cos x,
\\
\dot \alpha = -\frac12 (\omega_rp^2 - \Delta)A - \frac12 \gamma
\alpha + \exp[-(\tau-\frac{3}{2} \sigma_{\tau})^2/\sigma^2_{\tau}]B\cos x,
\\
\dot B = \frac12 (\omega_rp^2 + \Delta)\beta -
\exp[-(\tau-\frac{3}{2} \sigma_{\tau})^2/\sigma^2_{\tau}]\alpha \cos x,
\\
\dot \beta = -\frac12 (\omega_rp^2 + \Delta)B+
\exp[-(\tau-\frac{3}{2} \sigma_{\tau})^2/\sigma^2_{\tau}]A\cos x,
\end{gathered}
\label{mainsys}
\end{equation}
where $x\equiv k_f X$ and $p\equiv P/\hbar k_f$ are atomic center-of-mass position
and transversal momentum, respectively and dot denotes differentiation with respect
to the dimensionless time $\tau\equiv \Omega_0 t$. The recoil frequency,
$\omega_r\equiv\hbar k_f^2/m_a\Omega_0\ll 1$,
the atom-laser detuning, $\Delta\equiv(\omega_f-\omega_a)/\Omega_0$,
the decay rate $\gamma=\Gamma/\Omega_0$, and the characteristic interaction time,
$\sigma_{\tau}\equiv r\Omega_0/v_z$, are the control parameters.

%
\begin{figure}[!tpb]
\includegraphics[width=0.3\textwidth]{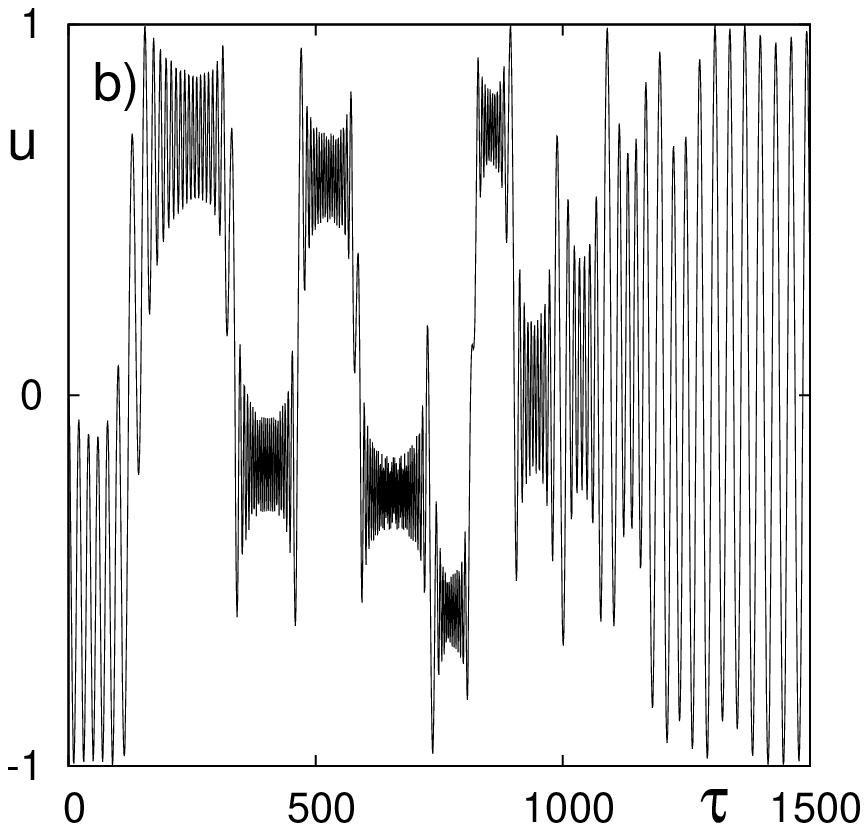}
\includegraphics[width=0.3\textwidth]{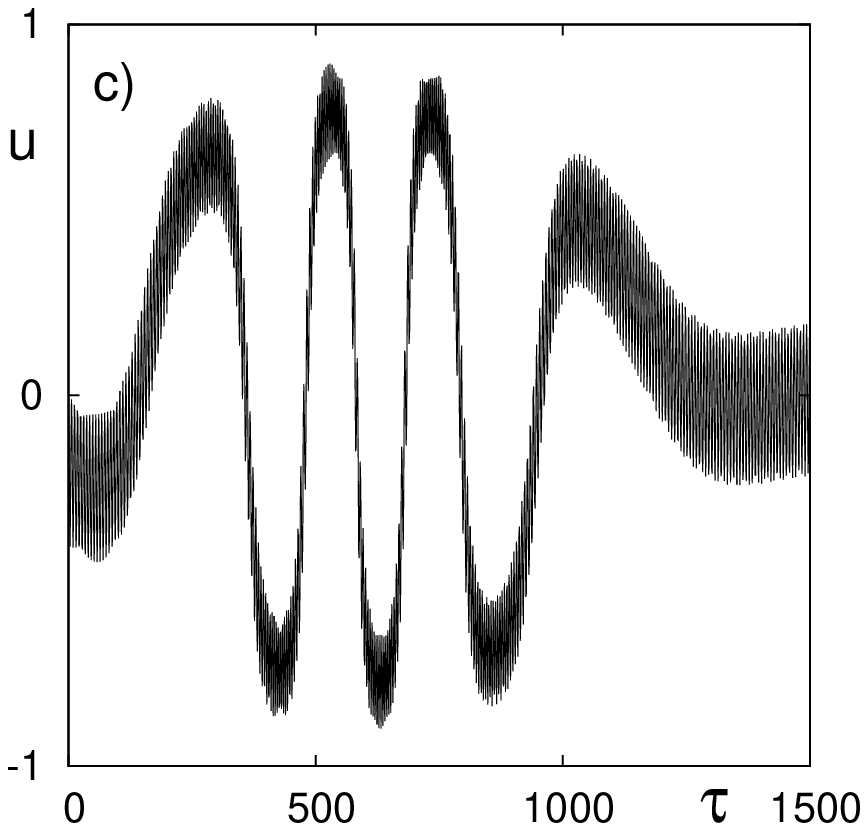}
\caption{Evolution ($\tau$ is in units of $\Omega_0^{-1}$) of the
atomic dipole-moment component $u=2(AB+\alpha\beta)$  in (a) the chaotic
($\Delta=0.2$) and (b) regular ($\Delta=1$) regimes of atomic motion.}
\label{fig3}
\end{figure}

Without the decay, $\gamma=0$, Eqs.~(\ref{mainsys}) constitute the nonlinear
Hamiltonian dynamical system with three degrees of freedom describing an
atom moving in the six-dimensional phase space. The simple way to know
how complicated this motion may be is to compute the
quantitative measure of chaos, maximal Lyapunov exponent
characterizing the mean rate of the
exponential divergence of initially close trajectories.
\revision {Because of a transient character of chaos we compute
the finite-time Lyapunov exponent $\lambda$, i.e. the value of
the exponent at the moment when atoms leave the interaction region.}
The result of computation with Eqs.~(\ref{mainsys})
at the given value of the recoil frequency,
$\omega_r=10^{-3}$, and zero decay rate in dependence on
the detuning $\Delta$ and the initial atomic transversal momentum
$p_0$ is shown in
Fig.~\ref{fig2}. Color codes the magnitude of \revision {the finite-time
Lyapunov exponent.} In white regions the values of $\lambda$ are almost zero,
and the internal and translational motion is regular in the corresponding ranges
of $\Delta$ and $p_0$. In shadowed regions positive values of $\lambda$ imply
unstable motion.
\begin{figure}[!tpb]
\includegraphics[width=0.3\textwidth]{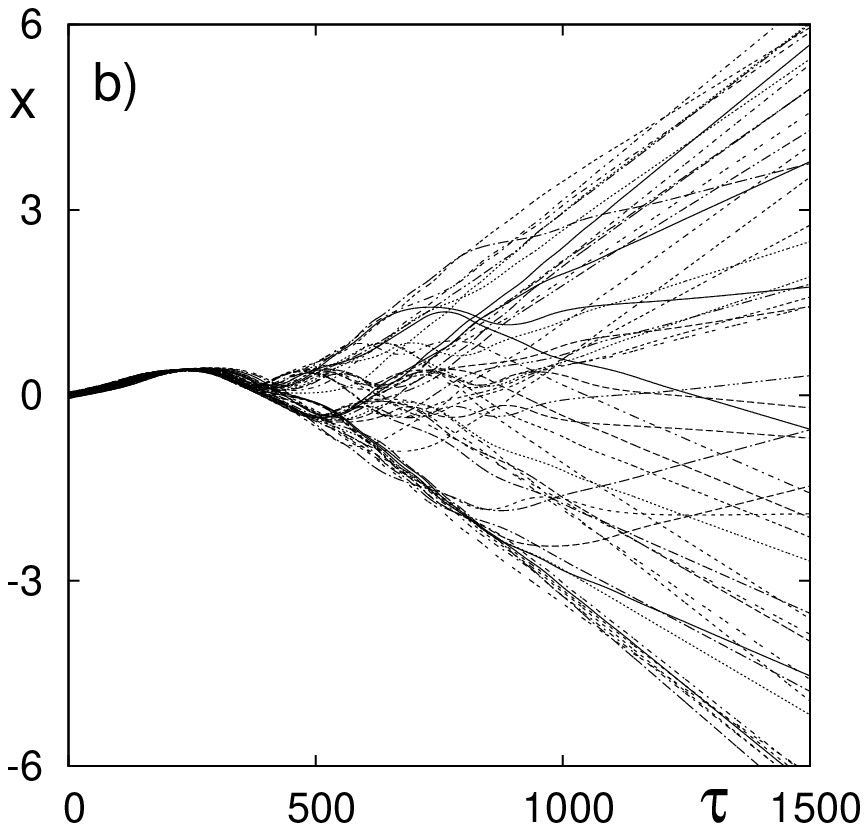}
\includegraphics[width=0.3\textwidth]{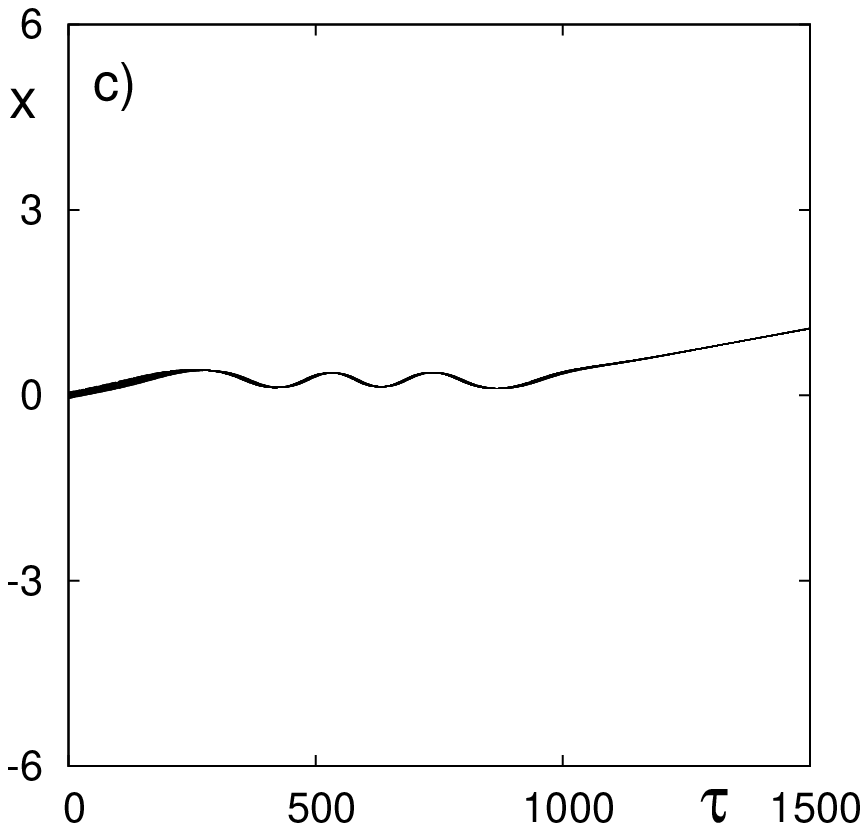}
\caption{Trajectories in the real space for 50 atoms without
spontaneous emission.
Hamiltonian (a) chaotic ($\Delta=0.2$) and (b) regular scattering
($\Delta=1$). The atomic position $x$ is in units of the optical wavelength.}
\label{fig4}
\end{figure}
\begin{figure}[!tpb]
\includegraphics[width=0.3\textwidth]{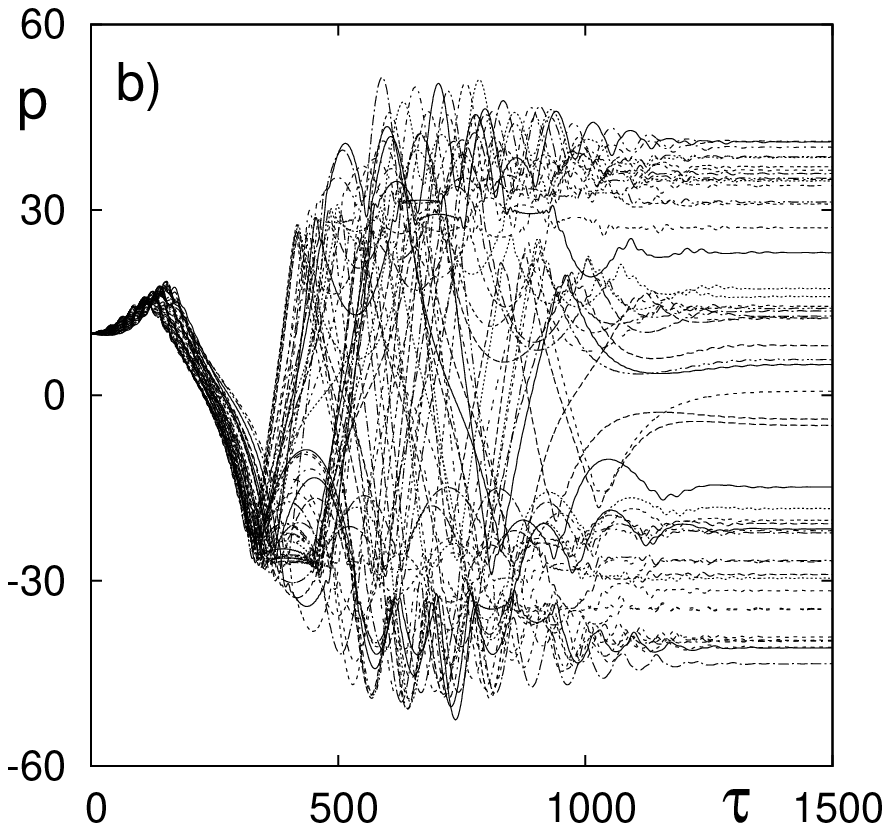}
\includegraphics[width=0.3\textwidth]{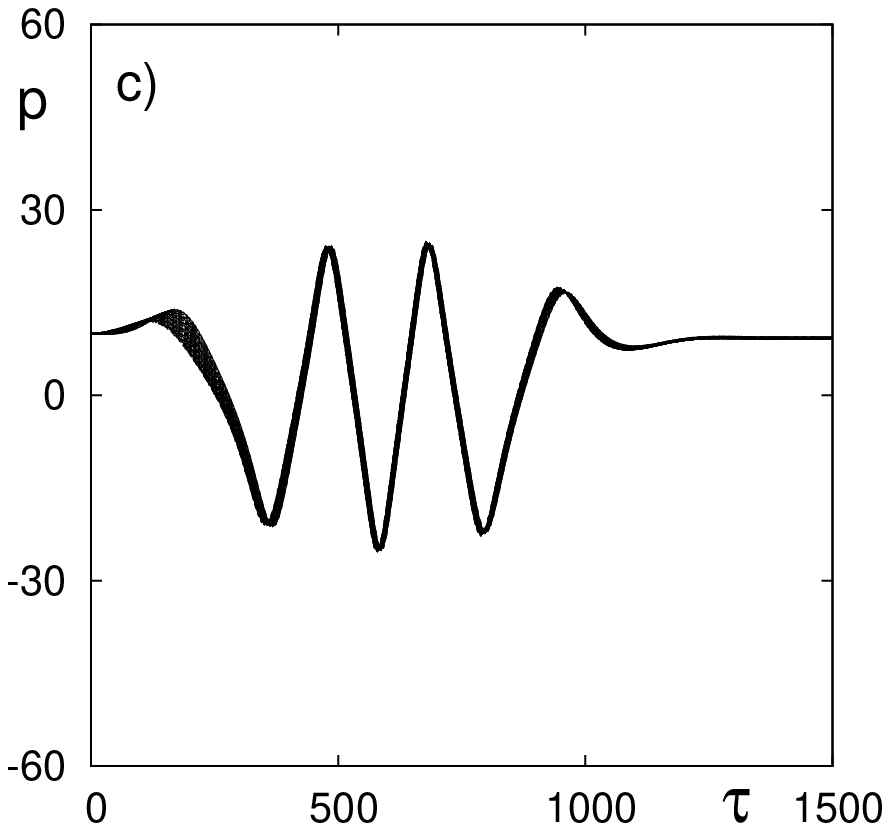}
\caption{The same as Fig.~\ref{fig4} but in the momentum
space.}
\label{fig5}
\end{figure}

The scheme of the proposed experiment in Fig.~\ref{fig1} resembles
the scattering problem with particles entering an interaction region along
completely regular trajectories and leaving it along asymptotically
regular trajectories \cite{PH86,PhysD,JETP04,Gaspard,E88,A09}. \revision {However,
in difference from the standard chaotic scattering with a  nonattractive
fractal invariant set existing over an infinite time, this process may be
interpreted as a transient chaos or a finite-time chaotic scattering.}

There are three possible chaotic regimes of the center-of-mass motion
along the SW optical axis \cite{PRA01,PRA07}. In dependence on the initial conditions
and the values of the control parameters, atoms may oscillate chaotically
in wells of the optical potential or move ballistically over its hills with
chaotic variations of their velocity. Chaotic motion
becomes possible in a narrow range of the detuning values, $0< |\Delta| <1$.
At $\Delta=0$, the synchronized electric-dipole component,
$u=2(AB+\alpha\beta)$ becomes a constant. That implies the additional integral
of motion in the Hamiltonian version of Eqs.(\ref{mainsys}) and
the regular motion with $\lambda=0$. Far from the resonance,
at $|\Delta| >1$, the motion is again regular both in the trapping
and flight modes.
That speculation is confirmed by the Lyapunov map in Fig.~\ref{fig2}.

Moreover, there is a specific type of motion, chaotic walking in
a deterministic optical potential, when atoms can change the direction
of motion alternating between flying through
the SW and being trapped in its potential wells.
We would like to stress that the local instability produces chaotic center-of-mass
motion in a rigid SW without any modulation of its
parameters in difference from the case with periodically
kicked and far detuned optical lattices \cite{GSZ92,R96,Steck01,C09}.
The trivial time dependence in the Hamiltonian (\ref{mainsys})
cannot produce chaotic motion, it simply accounts for a modulation of
the interaction of atoms with a Gaussian laser beam.
Even if the atoms would cross an absolutely homogeneous (in the
$z$-direction) laser beam there would be under appropriate conditions
chaotic atomic center-of-mass motion in the transversal $x$-direction.

Chaotic walking occurs due to the specific
behavior of the Bloch-vector component,
$u$, of a moving atom whose shallow oscillations between the SW nodes are interrupted by
sudden jumps with different amplitudes while atom crosses each
node~\cite{PRA07}. We illustrate in Fig.~\ref{fig3} the behavior of the $u$
component with different values of the detuning $\Delta=0.2$ and $\Delta=1$
at which the atomic motion in accordance with
the $\lambda$-map in Fig.~\ref{fig2} is chaotic
and regular, respectively. The time of the atomic interaction with the SW field
is estimated to be $3\sigma_{\tau}=1200$. So, the jumps of the $u$ variable
(if any) disappear after that time in Fig.~\ref{fig3}. It follows from the second
equation in the set ~(\ref{mainsys}) that jumps in the variable
$u=2(AB+\alpha\beta)$ result in jumps of the atomic momentum while crossing
a node of the SW. If the value of the atomic energy is close to a separatrix
one, the atom after the corresponding jump-like change in $p$ can either
overcome the potential barrier and leave a potential well or
it will be trapped by the well, or it will move as before.
The jump-like behavior of $u$ is the ultimate reason of chaotic atomic
walking along a deterministic SW.

It is easy to estimate the range of initial momenta at which atoms are
expected to change their direction of motion or move ballistically.
At small detunings, $|\Delta| \ll 1$,
the total energy consists of the kinetic one, $K=\omega_rp^2/2$,
and the potential one,  $U=  u\cos x$.
If $K(\tau=0)> |U_{\rm max}|=1$, then the atom will move
ballistically. This occurs if the initial atomic
momentum, $p_0$, satisfies to the condition $p_0> \sqrt{2/\omega_r}>44$.
If the initial conditions are chosen to give
$0\le K(\tau=0)+U(\tau=0) \le 1$, the  atoms with
$0 \lesssim p_0 \lesssim  44$ are expected to perform a chaotic walking
at positive $\Delta$.

\subsection{Hamiltonian chaotic scattering}
\begin{figure}[htb]
\includegraphics[width=0.3\textwidth]{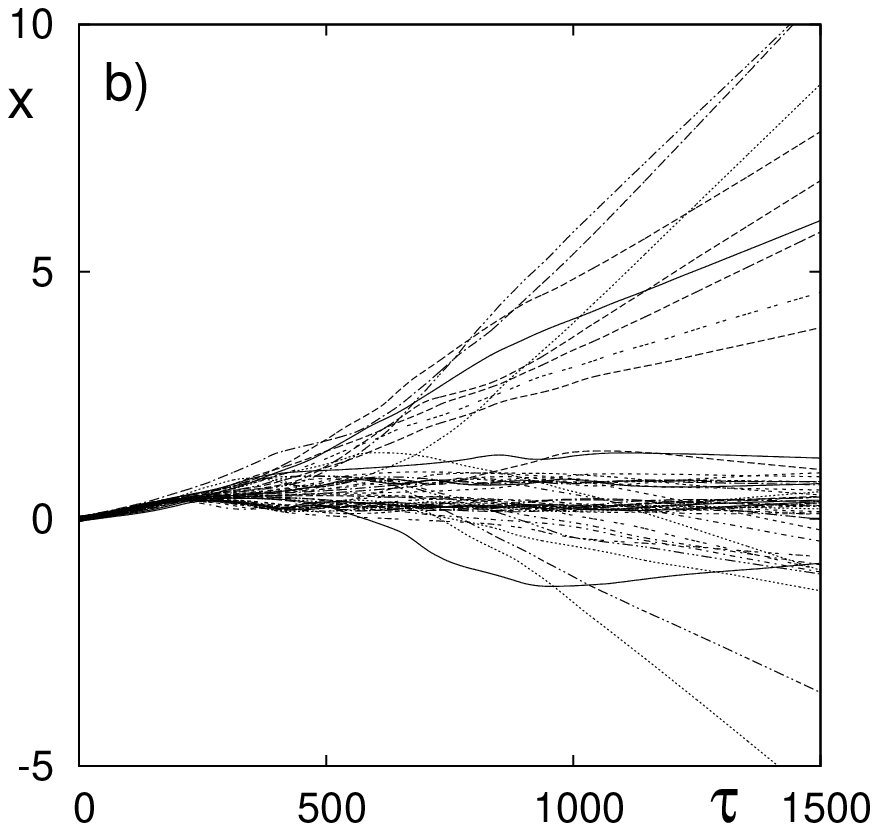}
\includegraphics[width=0.3\textwidth]{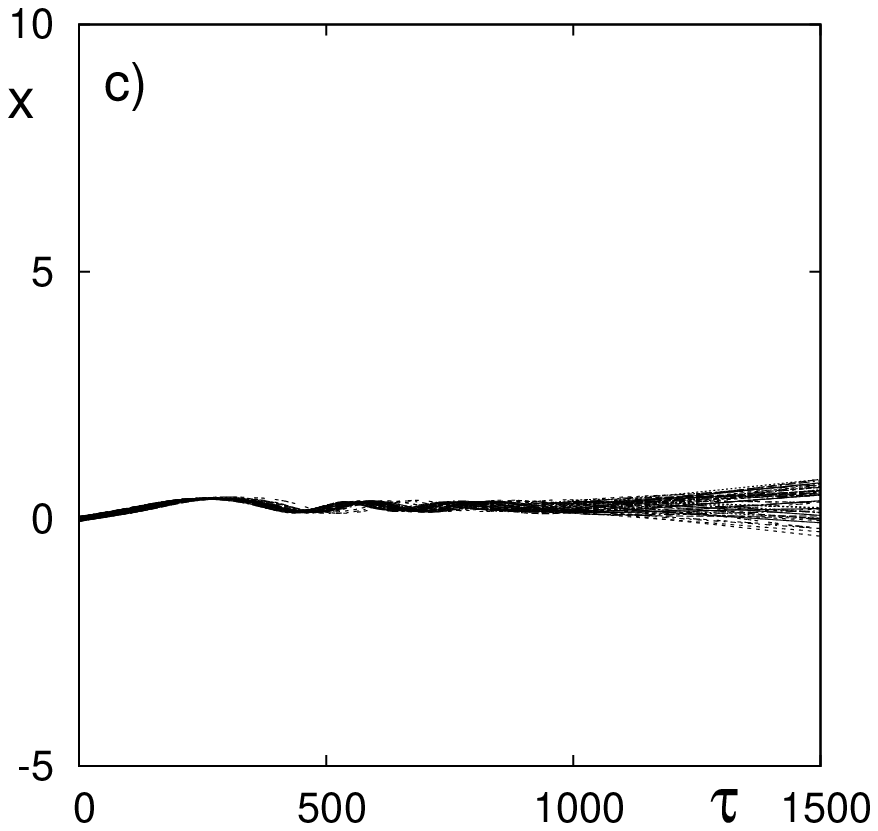}
\caption{Scattering of 50 spontaneously emitting atoms at the SW
with the decay rate $\gamma=0.05$ and the same other conditions as in
Fig.~\ref{fig4}.
(a) Chaotic  ($\Delta=0.2$) and (b) regular ($\Delta=1$) regimes.}
\label{fig6}
\end{figure}
\begin{figure}[htb]
\includegraphics[width=0.3\textwidth]{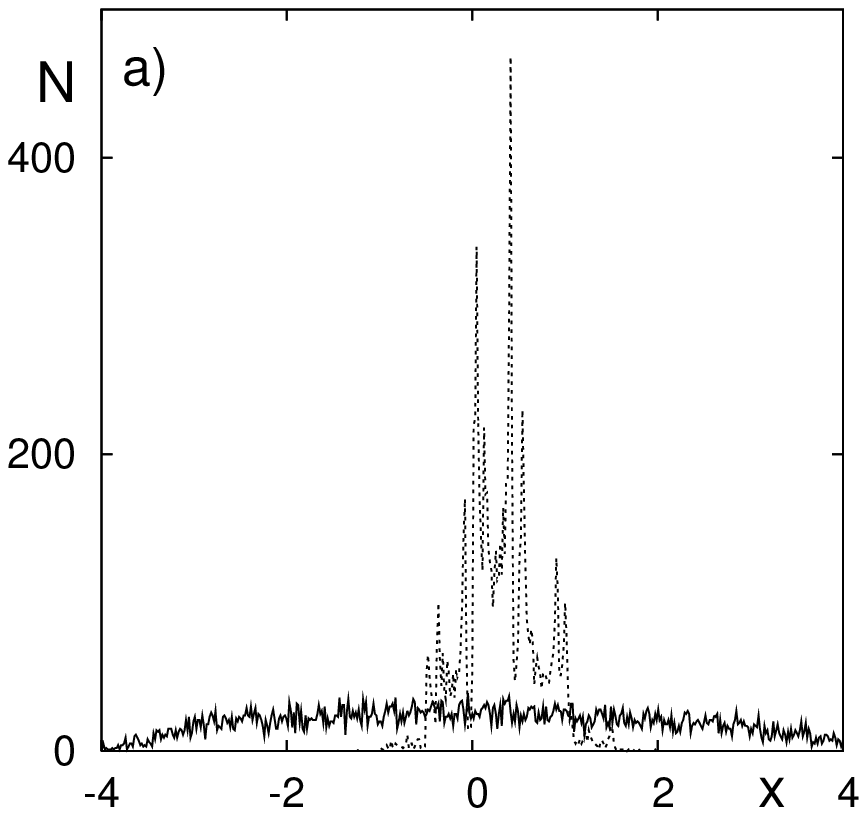}
\includegraphics[width=0.3\textwidth]{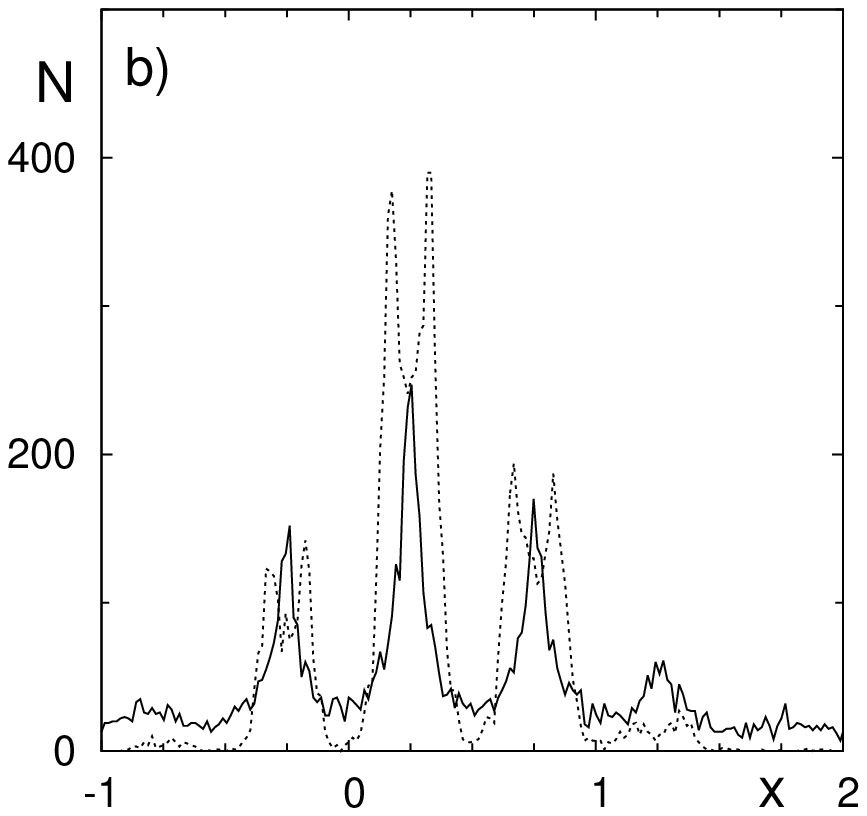}
\caption{The distributions of $10^4$ lithium atoms (a) without
and (b) with spontaneous emission at $\tau=1000$ under the conditions of chaotic scattering at $\Delta=0.2$
(bold curves) and regular scattering  at $\Delta=1$ (dashed curves).}
\label{fig7}
\end{figure}

Let us consider a spatially uniform
and previously focused beam of atoms crossing a Gaussiam laser beam.
The position and
momentum distributions of atoms are measured after interaction with the SW field.
We predict that those distributions would be much broader at those
values of $\Delta$ at which one expects chaotic walking to occur.
Firstly, we perform simulation with a negligible probability
of spontaneous emission. To be concrete let us take
calcium atoms with the working intercombination transition $4^1S_0-4^3P_1$
at $\lambda_a = 657.5$~nm, the recoil frequency $\nu_{\rm rec}\simeq 10$~KHz,
and the lifetime of the excited state $T_{\rm sp} = 0.4$~ms.
Taking the maximal Rabi frequency to be $\Omega_0/2\pi = 2\cdot 10^7$~Hz,
the radius of the laser beam $r = 0.3$~cm, and the mean longitudinal velocity
$v_z=10^3$~m/s, the interaction time is estimated to be $0.9$~ms.
The normalized recoil frequency is $\omega_r = 4\pi \nu_{\rm rec}/
\Omega_0=10^{-3}$ and $\sigma_{\tau}=400$.

We numerically solve the equations of motion (\ref{mainsys}) with $\gamma=0$
at two values
of $\Delta$ corresponding to  chaotic and regular
regimes of the center-of-mass motion.
In accordance with the Lyapunov map in Fig.~\ref{fig2}, behavior of the Bloch
component $u$ in Fig.~\ref{fig3}, and the simple estimates
given above, we expect the chaotic scattering  of atoms at $\Delta=0.2$
and their
regular motion at $\Delta=1$. Trajectories in the real and
momentum spaces for 50 atoms with the same initial momentum,
$p_0=10$, and initial positions in the range $-\pi/10\le x \le\pi/10$
are shown in Figs.~\ref{fig4} and 5, respectively, at the fixed
value of the recoil frequency $\omega_r=10^{-3}$.
The upper panels in Figs.~\ref{fig4} and 5 illustrate the broad distributions
of the atoms in the $x$
and $p$ spaces in the regime of chaotic scattering
that contrasts strictly
with those obtained in the case of the regular scattering
at $\Delta=1$ (Figs.~\ref{fig4}b and 5b). In order to create narrow
atomic beams, one may use a pair of
light masks. The first SW with a red large detuning ($\Delta <0$)
splits the initial atomic beam into a number of narrow beams
with the widths much smaller than the optical wavelength which then
cross the second SW. \revision {The method for creating narrow wave packets
in the nonadiabatic regime of scattering has been proposed in
Ref.\cite{Fedorov}.}

\subsection{Dissipative chaotic scattering and simulation of a real experiment}

We have illustrated in Figs.~\ref{fig4} and 5 the Hamiltonian chaotic scattering
that may occur in the
absence of any losses. To simulate trajectories of spontaneously emitting atoms
we use the standard stochastic wave-function technique (see, for example,
\cite{Carmichael,Dal,Dum})
for solving Eqs.~(\ref{mainsys}).
The integration time is divided into a large number of small time intervals
$\delta\tau$. At the end of the first one $\tau=\tau_1$ the
probability of spontaneous emission,
$s_1=\gamma \delta \tau |a_{\tau_1}|^2/(|a_{\tau_1}|^2 + |b_{\tau_1}|^2)$,
is computed and compared with a random number $\varepsilon$ from the interval
$[0,1]$. If $s_1 < \varepsilon_1$, then one prolongs the integration
but renormalizes the state vector in the end of the first interval
at $\tau= \tau_1^+$: $a_{\tau_1^+}=a_{\tau_1}/\sqrt{|a_{\tau_1}|^2
+ |b_{\tau_1}|^2}$ and $b_{\tau_1^+}=b_{\tau_1}/\sqrt{|a_{\tau_1}|^2 +
|b_{\tau_1}|^2}$. If $s_1 \ge \varepsilon_1$, then the atom emits a
spontaneous photon and performs the jump to the ground state at $\tau= \tau_1$
with $A_{\tau_1}=\alpha_{\tau_1}=\beta_{\tau_1}=0$, $B_{\tau_1}=1$. Its momentum
in the $x$ direction changes for a random number from the interval
$[0,1]$ due to the photon recoil effect, and the next time step commences.

We simulate lithium atoms with the relevant
transition $2S_{1/2}-2P_{3/2}$, the corresponding wavelength
$\lambda_a = 670.7$~nm, recoil frequency $\nu_{\rm rec}=63$~KHz,
and the decay time $T_{\rm sp} = 2.73 \cdot 10^{-8}$~s.
With the maximal Rabi frequency $\Omega_0/2\pi \simeq 126$~MHz
and the radius of the laser beam $r = 0.05$~cm
one gets $\omega_r = 10^{-3}$, $\sigma_{\tau}=400$, and $\gamma=0.05$.
Simulated trajectories in the real space for 50 spontaneously emitting atoms
under the same conditions
as in Fig.~\ref{fig4} are shown in Fig.~\ref{fig6}.
Even though deterministic
Hamiltonian chaos is masked by random events of spontaneous emission, nevertheless
the spatial and momentum (not shown) distributions are much broader at those values of
$\Delta$ at which the Hamiltonian center-of-mass motion is chaotic. Namely
the chaotic Hamiltonian walking is eventually responsible for divergency of
atomic beams in the real and momentum spaces.

To simulate a real experiment we consider a beam of $10^4$ lithium atoms with
the initial Gaussian distribution
(the rms $\sigma_x=\sigma_p=2$ and the average values, $x_0=0$,
and momentum, $p_0=10$) and compute their distribution at a fixed moment of time.
In Fig.~\ref{fig7}a we compare the atomic position distributions
at $\tau=1000$ for the chaotic
scattering at $\Delta =0.2$ (bold curve) and the regular scattering at
$\Delta =1$ (dashed curve) when neglecting spontaneous emission.
The difference is evident. In the regime of the chaotic scattering at
$\Delta =0.2$ atoms are distributed more or less homogeneously over a large
distance of 8 wavelengths along the $x$-axis whereas they form a few peaks
in a much more narrow interval under the conditions of the regular scattering at
$\Delta =1$. Figure~\ref{fig7}b demonstrates the distributions
of spontaneously emitting atoms at the normalized decay rate $\gamma=0.05$ under
the same conditions as in Fig.~\ref{fig7}a. The regularly scattered atoms at
$\Delta =1$ (dashed curve) form the contrast atomic relief with the bifurcated
peaks around the first few SW nodes at $x=\pm 1/4$, $x=\pm 3/4$ and $x= 5/4$.
The distribution of chaotically scattered atoms at $\Delta =0.2$ (bold curve)
has the peaks without any bifurcation at $x=\pm 1/4$ and $x=3/4$ with a
smaller number of atoms in each one. Moreover, this distribution is less
contrast as compared to the previous one.
Thus, we predict that under the conditions of chaotic scattering
there should appear less contrast and more broadened atomic reliefs as
compared to the case of regular
scattering because a large number of atoms are expected to be deposited
between the nodes as a result of chaotic walking along the SW axis.
The effect is expected to be more prominent under the coherent evolution
but it seems to be observable with spontaneously emitting atoms as well.

We predict that experiments on the scattering of atomic beams at a SW laser
field can directly image chaotic walking of atoms along the SW.
In a real
experiment the final spatial distribution can be recorded via fluorescence or
absorption imaging on a CCD, commonly used methods in atom optics experiments
yielding information on the number of atoms and the cloud's spatial size.
The other possibility is a nanofabrication
where the atoms after the interaction with the SW are deposited on a silicon
substrate in a high vacuum chamber. In this case the spatial distribution
can be analyzed with an atomic force microscope. As to the momentum
distribution, it can be measured, for example, by a time-of-flight
technique. The modern tools of atom optics enable to create
narrow initial atomic distributions in position and momentum, reduce coupling
to the environment and technical noise, create one-dimensional optical
potentials, and to measure spatial and momentum distributions
with high sensitivity and accuracy \cite{Steck01}.

\section{Conclusion}

We have simulated the new type
of atomic diffraction at a SW without any modulation of its parameters and
shown that it can be observed in real experiments. That would
be the prove of existence of the novel
type of atomic motion, chaotic walking in a deterministic environment.
The effect could be used in optical nanolithography to fabricate complex
atomic structures on substrates.

\section{Acknowledgments}
The work was supported
by the Integration grant from the Far-Eastern
and Siberian branches of the Russian Academy of Sciences (12-II-0-02-001),
and by the Program ``Fundamental Problems of  Nonlinear Dynamics in Mathematics
and Physics''. I thank L.E. Konkov and M.Yu. Uleysky for the help in preparing some figures.

\end{document}